\documentclass[11pt]{article}
\usepackage{a4p}
\usepackage{epsfig}
\usepackage{array}
\usepackage{cite}
\usepackage{pennames}
\usepackage{rotating}
\usepackage{comment}
\includecomment{paper}
\newcommand{\PPEnum} {CERN-EP-2000-017}
\newcommand{\Date}      {21 Jan 2000}

\newcommand{\mybold} {\boldmath }

\def\lumia{55}
\def\lumib{178}
\def\emna{182.62}
\def\emnb{188.63}
\def\xsca{ 
 0.12 \ ^{+ 0.20 }_{-0.18 } (\mbox{stat.}) 
      \ ^{+ 0.03 }_{-0.02 } (\mbox{syst.})
}
\def\xsclma{0.55}
\def\xscb{ 
 0.80 \ ^{+ 0.14 }_{-0.13 } (\mbox{stat.}) 
      \ ^{+ 0.06 }_{-0.05 } (\mbox{syst.})}
\def\bbrb{ 0.21 ^{+0.07}_{-0.06} (\mathrm{stat.})
                   \pm  0.01 (\mathrm{syst.})  }
\def\bbxs{0.75 ^{+0.15 }_{-0.14 } (\mbox{stat.}) 
                 \ ^{+ 0.06}_{-0.05 } (\mbox{syst.})}
\def\meqqeea{(55.2 \pm 2.7)}
\def\meqqmma{(66.9 \pm 2.6)}
\def\meqqtta{(22.2 \pm 1.7)}

\def\meqqnna{(30.5 \pm 2.0)}
\def\meqqqqa{(25.3 \pm 1.3)}
\def\meqqbba{(35.1 \pm 2.4)}

\def\meqqeeb{(65.1 \pm 2.9)}
\def\meqqmmb{(72.3 \pm 2.8)}
\def\meqqttb{(26.9 \pm 2.0)}

\def\meqqnnb{(33.9 \pm 2.3)}
\def\meqqqqb{(33.0 \pm 1.8)}
\def\meqqbbb{(38.6 \pm 2.7)}

 \newcommand{\lllllb} {a}
 \newcommand{\eennlb} {b}
 \newcommand{\mmnnlb} {c}
 \newcommand{\qqeelb} {d}
 \newcommand{\bbeelb} {e}
 \newcommand{\qqmmlb} {f}
 \newcommand{\bbmmlb} {g}
 \newcommand{\qqttlb} {h}
 \newcommand{\bbttlb} {i}
 \newcommand{\xbttlb} {j}
 \newcommand{\qqnnlb} {k}
 \newcommand{\bbnnlb} {l}
 \newcommand{\qqqqlb} {m}
 \newcommand{\qqxblb} {n}
 \newcommand{\qqbblb} {o}
\newcommand{\inmath}[1] {\ifmmode#1\else$#1$\fi}
\newcommand{\definmath}[2] {\def#1{\ifmmode#2\else$#2$\fi}}
\newcommand{\gVe}  {g^{\mathrm{e}}_{\mathrm{V}}}
\newcommand{\gAe}  {g^{\mathrm{e}}_{\mathrm{A}}}

\newcommand{\PZ}   {\mbox{$\mathrm{Z}$}}  
\definmath{\PWpm} {\mathrm{W}^{\pm}}      
\definmath{\Plp} {\ell^{+}}        
\definmath{\Plm} {\ell^{-}}        
\definmath{\Plpm}   {\ell^{\pm}}         
\definmath{\Pgtp} {\tau^{+}}        
\definmath{\Pgtm} {\tau^{-}}        
\definmath{\Pgtpm}   {\tau^{\pm}}         
\definmath{\Pgn}  {\nu}          
\definmath{\Pagn} {\overline{\nu}}     
\definmath{\Pf}      {\mathrm{f}}
\definmath{\Paf}  {\overline{\mathrm{f}}}
\definmath{\Pq}      {\mathrm{q}}
\definmath{\Paq}  {\overline{\mathrm{q}}}
\definmath{\Pu}      {\mathrm{u}}
\definmath{\Pau}  {\overline{\mathrm{u}}}
\definmath{\Pd}      {\mathrm{d}}
\definmath{\Pad}  {\overline{\mathrm{d}}}
\definmath{\Ps}      {\mathrm{s}}
\definmath{\Pas}  {\overline{\mathrm{s}}}
\definmath{\Pc}      {\mathrm{c}}
\definmath{\Pac}  {\overline{\mathrm{c}}}
\definmath{\Pb}      {\mathrm{b}}
\definmath{\Pab}  {\overline{\mathrm{b}}}
\definmath{\Pt}      {\mathrm{t}}
\definmath{\Pat}  {\overline{\mathrm{t}}}
\definmath{\Pap}  {\overline{\mathrm{p}}}
\definmath{\Pan}  {\overline{\mathrm{n}}}
\definmath{\PaD}  {\overline{\mathrm{D}}}
\definmath{\PaDz} {\overline{\mathrm{D}}^{0}}
\definmath{\PaB}  {\overline{\mathrm{B}}}
\definmath{\PaBz} {\overline{\mathrm{B}}^{0}}
\definmath{\PsDpm}   {\mathrm{D}^{\pm}_{\mathrm{s}}}  
\definmath{\PcgLpm}  {\Lambda^{\pm}_{\mathrm{c}}}  
\definmath{\PD} {\mathrm{D}}     
\definmath{\PDst} {\mathrm{D}^{*}}     
\definmath{\PgLz} {\Lambda^{0}}        

\newcommand{\massof}[1] {m_{\smash{#1}\mathstrut}}

\newcommand{\mPZ} {\massof{\mathrm{Z}}}

\newcommand{\mtautau}  {\massof{\mathrm{\tau\tau }}}
\newcommand{\GammaZ} {\Gamma_{\mathrm{Z}}}

\newcommand{\epem}   {\Pep\Pem}

\newcommand{\mumu}   {\Pgmp\Pgmm}
\newcommand{\tautau} {\Pgtp\Pgtm}

\newcommand{\ellell} {{\Plp}{\Plm}}
\newcommand{\nunu}   {\Pgn\Pagn}
\newcommand{\qqbar}  {\Pq\Paq}

\newcommand{\bbbar}  {\Pb\Pab}

\newcommand{\zz}           {\PZ \PZ}
\newcommand{\eetozz}           {\epem \to \zz}
\newcommand{\eetozg}           {\epem \to \PZ \gamma^*}
\newcommand{\zztollll}         {{ \zz \to {\ellell} {\ellell}}}
\newcommand{\zztoqqll}         {{ \zz \to  \qqbar \ellell}}
\newcommand{\zztoqqee}         {{ \zz \to  \qqbar \epem}}
\newcommand{\zztoqqmm}         {{ \zz \to  \qqbar \mu^+ \mu^- }}

\newcommand{\zztobbll}         {{ \zz \to  \bbbar\ellell}}
\newcommand{\zztoqqnn}         {{ \zz \to  \qqbar \nunu}}
\newcommand{\zztobbnn}         {{ \zz \to  \bbbar \nunu}}
\newcommand{\zztollnn}         {{ \zz \to \ellell \nunu}}
\newcommand{\zztoqqqq}         {{ \zz \to \qqbar \qqbar}}
\newcommand{\zztoqqbb}         {{ \zz \to \qqbar \bbbar}}

\newcommand{\nnnn}       { \nunu \nunu}
\newcommand{\llnn}       { \ellell \nunu}
\newcommand{\eenn}       {\epem \nunu}
\newcommand{\mmnn}       {\mu^+ \mu^- \nunu}
\newcommand{\ttnn}       {\tau^+ \tau^- \nunu}
\newcommand{\llll}       {  \ellell \ellell}
\newcommand{\eeee}       { \epem \epem}
\newcommand{\eemm}       { \epem \mu^+ \mu^-}
\newcommand{\eett}       { \epem \tau^+ \tau^-}
\newcommand{\mmmm}       { \mu^+ \mu^-  \mu^+ \mu^-}
\newcommand{\mmtt}       { \mu^+ \mu^- \tau^+ \tau^-}
\newcommand{\tttt}       { \tau^+ \tau^- \tau^+ \tau^-}
\newcommand{\qqlnu}      { \mathrm{\qqbar} \ell \nu}
\newcommand{\qqll}       {  \qqbar \ellell}
\newcommand{\bbll}       { \bbbar \ellell}
\newcommand{\qqnn}       { \qqbar \nunu}
\newcommand{\bbnn}       { \bbbar \nunu}
\newcommand{\qqqq}       { \qqbar \qqbar}
\newcommand{\qqbb}       { \qqbar \bbbar}
\newcommand{\qqee}       { \qqbar \epem}
\newcommand{\bbee}        {\bbbar \epem}
\newcommand{\qqmm}       { \qqbar \mu^{+} \mu^{-} }
\newcommand{\bbmm}       { \bbbar \mu^{+} \mu^{-} }
\newcommand{\qqtt}       { \qqbar \tau^{+} \tau^{-}}
\newcommand{\bbtt}       { \bbbar  \tau^{+} \tau^{-}}
\newcommand{\llllz}      {\llll}
\newcommand{\qqeez}      {\qqee \, \& \, \overline{\bbee}}
\newcommand{\bbeez}      {\qqee \, \& \, \bbee      }
\newcommand{\qqmmz}      {\qqmm \, \& \, \overline{\bbmm}}
\newcommand{\bbmmz}      {\qqmm \, \& \, \bbmm      }
\newcommand{\qqttz}      {\qqtt \, \& \,  \overline{\bbtt}}
\newcommand{\bbttz}      {\qqtt \, \& \,  \bbtt      }
\newcommand{\xbttz}      {\bbtt \, \& \,  \overline{\qqtt}}
\newcommand{\qqnnz}      {\qqnn \, \& \,  \overline{\bbnn}}
\newcommand{\bbnnz}      {\qqnn \, \& \,  \bbnn      }

\newcommand{\qqqqz}      {\qqqq \, \& \,  \overline{\qqbb}}
\newcommand{\qqxbz}      {\qqbb \, \& \,  \overline{\qqqq}}
\newcommand{\qqbbz}      {\qqbb \, \& \,  \qqqq      }
\newcommand{\eennz}      {\eenn}
\newcommand{\mmnnz}      {\mmnn}
\newcommand{\ZZZ}[1]      { f_{#1}^{\mathrm{ZZZ}} }
\newcommand{\ZZG}[1]      { f_{#1}^{\mathrm{ZZ\gamma}} }
\newcommand{\ZZZR}[1]      { Re \{ f_{#1}^{\mathrm{ZZZ}} \} }
\newcommand{\ZZGR}[1]      { Re \{ f_{#1}^{\mathrm{ZZ\gamma}} \} }
\newcommand{\ZZZI}[1]      { Im \{ f_{#1}^{\mathrm{ZZZ}} \} }
\newcommand{\ZZGI}[1]      { Im \{ f_{#1}^{\mathrm{ZZ\gamma}} \} }

\newcommand{\roots} {\sqrt{s}}
\newcommand{\rootsp} {\sqrt{s'}}

\newcommand{\Ebeam}  {E_{\mathrm{b}}}
\newcommand{\Evis}   {\mbox{$E_{\mathrm{vis}}$}}

\definmath{\GeV}  {\mathrm{GeV}}
\definmath{\GeVc} {\mathrm{GeV}\!/c}
\definmath{\GeVcc}   {\mathrm{GeV}\!/c^2}
\definmath{\MeV}  {\mathrm{MeV}}
\definmath{\MeVc} {\mathrm{MeV}\!/c}
\definmath{\MeVcc}   {\mathrm{MeV}\!/c^2}
\definmath{\MVm}  {\mathrm{MV}\!/\mathrm{m}}
\definmath{\keV}  {\mathrm{keV}}
\definmath{\keVcm}   {\mathrm{keV}\!/\mathrm{cm}}
\definmath{\kV}      {\mathrm{kV}}
\definmath{\km}      {\mathrm{km}}
\definmath{\meter}   {\mathrm{m}}
\definmath{\cm}      {\mathrm{cm}}
\definmath{\mm}      {\mathrm{mm}}
\definmath{\micron}  {\mu\mathrm{m}}
\definmath{\nm}      {\mathrm{nm}}
\definmath{\kg}      {\mathrm{kg}}
\definmath{\gram} {\mathrm{g}}
\definmath{\second}  {\mathrm{s}}
\definmath{\microsec}   {\mu\mathrm{s}}
\definmath{\degree}  {^\circ}
\definmath{\degC} {^\circ\mathrm{C}}
\definmath{\ohm}  {\Omega}
\definmath{\Mohm} {\mathrm{M}\Omega}
\definmath{\rad}  {\mathrm{rad}}
\definmath{\mrad} {\mathrm{mrad}}
\definmath{\nb}      {\mathrm{nb}}
\definmath{\pb}      {\mathrm{pb}}
\newcommand{\eqref}[1]  {(\ref{#1})}
\newcommand{\PhysLett}  {Phys.~Lett.}

\newcommand{\PhysRev}   {Phys.~Rev.}
\newcommand{\NPhys}  {Nucl.~Phys.}
\newcommand{\NIM} {Nucl.~Instrum.\ Methods}

\newcommand{\IEEENS} {IEEE Trans.\ Nucl.~Sci.}
\newcommand{\CPC} {Comput. Phys. Commun.}
\newcommand{\EPJ} {Eur.~Phys.~J.} 
\newcommand{\OPALColl}    {OPAL Collab.}
\newcolumntype{L} {>{$}l<{$}}
\newcolumntype{C} {>{$}c<{$}}
\newcolumntype{R} {>{$}r<{$}}

\newcommand{\mee}{m_{\rm ee}}

\newcommand{\mmumu}{m_{\mu\mu}}

\newcommand{\mqq}    {m_{\mathrm{qq}}}

\newcommand{\pt}     {p_{\mathrm{t}}}
\newcommand{\pti}    {p_{\mathrm{t}i}}
\newcommand{\ptj}    {p_{\mathrm{t}j}}

\newcommand{\mvis}   {m_{\mathrm{vis}}}
\newcommand{\mll}    {m_{\ell \ell}}
\newcommand{\mrec}   {m_{\mathrm{recoil}}}

\newcommand{\LWW}    { {\cal L}_{\PW \PW}}
\newcommand{\Lqqnn}  { {\cal L}_{\qqnn}}

\newcommand{\Wenu}   { \PW  \Pe \nu}
\newcommand{\Leenn}  { {\cal L}_{\eenn}}
\newcommand{\Lmmnn}  { {\cal L}_{\mmnn}}
\newcommand{\xsec}  {\sigma_{\mathrm{ZZ}}}
\newcommand{\nexp}  {\mu_{\mathrm{e}}}
\newcommand{\nobs}  {n_{\mathrm{obs}}}
\newcommand{\nback} {n_{\mathrm{back}}}
\newcommand{\nsm}   {n_{\mathrm{ZZ}}}
\newcommand{\smtot} {n_{\mathrm{SM}}}
\newcommand{\eff}   {\epsilon_{\mathrm{chan}}}
\newcommand{\br}    {  B_{\mathrm{ZZ} }}
\newcommand{\lint}  { L_{\mathrm{int}} }
\begin{document}
%
%
\begin{titlepage}
%
\begin{center}
    \Large EUROPEAN ORGANIZATION FOR NUCLEAR RESEARCH
\end{center}
\bigskip 
\begin{flushright}
    \large \PPEnum\\
    \Date \\
\end{flushright}
%
%
\begin{center}
    \huge\bf\boldmath
Z Boson Pair Production
in $\epem$ Collisions at $\sqrt{s}$=183 and 189\,GeV 
\end{center}\bigskip\bigskip
\begin{center}{\LARGE The OPAL Collaboration
}\end{center}\bigskip\bigskip
\bigskip\begin{center}{\Large  Abstract}\end{center}
%
%
A study of $\PZ$ boson pair production 
in $\epem$ annihilation at center-of-mass
energies near
$183$\,GeV and  $189$\,GeV is reported.
Final states containing only leptons,
($\llll$ and $\llnn$),
quark and lepton pairs,
($\qqll$, $\qqnn$)
and the all-hadronic final state ($\qqqq$) are considered.
In all states with at least one $\PZ$ boson decaying
hadronically, $\qqbar$ and $\bbbar$ final states
are  considered separately using lifetime and event-shape tags,
thereby improving the cross-section measurement.
At \mbox{$\roots = 189$\,GeV} the $\PZ$-pair cross section was measured to be
$\xscb \ \pb$, consistent with the Standard Model prediction.
At \mbox{$\roots = 183$\,GeV} the 95\% C.L. upper limit
is $\xsclma\ \pb$.
Limits on  anomalous ZZ$\gamma$ and ZZZ couplings are
derived.
\bigskip\bigskip\bigskip\bigskip
\bigskip\bigskip
\begin{center}
{\large Submitted to \PhysLett\ B}
\end{center}
%
%
\end{titlepage}
\begin{center}{\Large        The OPAL Collaboration
}\end{center}\bigskip
\begin{center}{
G.\thinspace Abbiendi$^{  2}$,
K.\thinspace Ackerstaff$^{  8}$,
P.F.\thinspace Akesson$^{  3}$,
G.\thinspace Alexander$^{ 22}$,
J.\thinspace Allison$^{ 16}$,
K.J.\thinspace Anderson$^{  9}$,
S.\thinspace Arcelli$^{ 17}$,
S.\thinspace Asai$^{ 23}$,
S.F.\thinspace Ashby$^{  1}$,
D.\thinspace Axen$^{ 27}$
% ...
}\end{center}\bigskip
\bigskip
$^{  1}$School of Physics and Astronomy, University of Birmingham,
Birmingham B15 2TT, UK
\newline
$^{  2}$Dipartimento di Fisica dell' Universit\`a di Bologna and INFN,
I-40126 Bologna, Italy
\newline
$^{  3}$Physikalisches Institut, Universit\"at Bonn,
D-53115 Bonn, Germany
\newline
% ...
$^{  8}$CERN, European Organisation for Particle Physics,
CH-1211 Geneva 23, Switzerland
\newline
$^{  9}$Enrico Fermi Institute and Department of Physics,
University of Chicago, Chicago IL 60637, USA
\newline
% ...
$^{ 16}$Department of Physics, Schuster Laboratory, The University,
Manchester M13 9PL, UK
\newline
$^{ 17}$Department of Physics, University of Maryland,
College Park, MD 20742, USA
\newline
% ...
$^{ 22}$Department of Physics and Astronomy, Tel Aviv University,
Tel Aviv 69978, Israel
\newline
$^{ 23}$International Centre for Elementary Particle Physics and
Department of Physics, University of Tokyo, Tokyo 113-0033, and
Kobe University, Kobe 657-8501, Japan
\newline
% ...
$^{ 27}$University of British Columbia, Department of Physics,
Vancouver BC V6T 1Z1, Canada
\newline
% ...

 
\section{Introduction}           \label{sec:intro}
\begin{figure}[b]
\center{
\epsfig{file=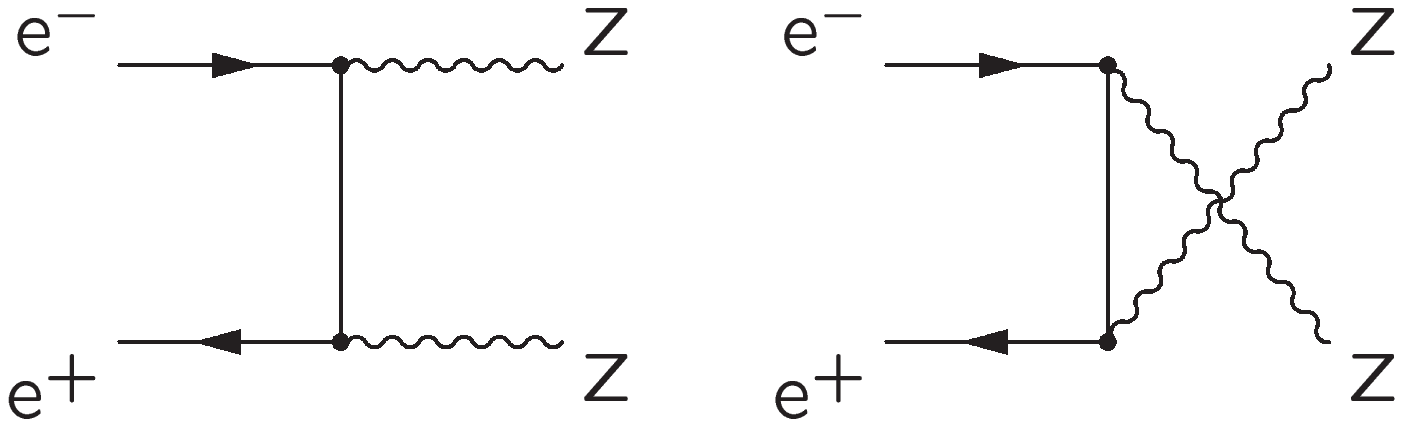,width=0.8\textwidth,
bbllx=40,bburx=595,bblly=400,bbury=650} }
\caption[$\eetozz$ ]{
\label{fig:eetozz}
NC2 Feynman diagrams for the process $\eetozz$ leading to
a final state with four fermions.
}
\end{figure}

The study of the process $\eetozz$ has recently become possible
since LEP now operates at center-of-mass energies 
above the threshold for on-shell $\PZ$ boson pair production.
In the Standard Model, the process $\eetozz$ occurs
via the NC2 diagrams~\cite{bib:zzfirst} shown in Figure~\ref{fig:eetozz}. 
The $\PZ$-pair cross section depends
on properties of the $\PZ$ boson ($\mPZ$, $\GammaZ$ and 
the vector and axial vector coupling of the $\PZ$ to 
electrons, $\gVe$ and $\gAe$) that have been measured with
great precision at the $\PZ$ resonance~\cite{bib:pdg}. 
The expected $\PZ$-pair cross section increases from about 
0.25~pb at \mbox{$\roots = 183$\,GeV} to about 1.0~pb at 
\mbox{$\roots = 200$\,GeV}, but remains more than an order of
magnitude smaller than $\PW$-pair production.
In contrast to $\PW$-pair production, where tree
level  $\PW\PW\gamma$ and $\PW\PW\PZ$ couplings are important, 
no  $\PZ\PZ\PZ$ and $\PZ\PZ\gamma$  couplings
are expected in the Standard Model. 
However, physics beyond the Standard Model could
lead to effective couplings~\cite{bib:hagiwara} 
which could then be observed
as deviations in the measured $\PZ$-pair cross section from the
Standard Model prediction.
Such deviations have been proposed
in the context of Higgs doublet models~\cite{bib:pal}
and in low scale gravity theories~\cite{bib:desh}.
In this paper we report on measurements of the NC2 $\PZ$-pair
cross section, including the extrapolation to final states with one
or both $\PZ$ bosons off-shell.  
These measurements, along with the angular distribution
of the observed events, are then used to extract limits
on possible $\PZ\PZ\PZ$ and $\PZ\PZ\gamma$ couplings.

%

In Section~\ref{sec:data} we describe the data
sets used and the Monte Carlo simulation
of signal and background.
In Section~\ref{sec:sel} we describe
the selection of the processes
$\zztollll$, $\zztollnn$, 
$\zztoqqll$, $\zztoqqnn$,  and $\zztoqqqq$,
where $\ellell$ denotes a charged lepton pair of opposite 
charge and $\qqbar$ any of the five lightest quark-antiquark pairs.
We also describe analyses of $\zztobbll$,
$\zztobbnn$ and $\zztoqqbb$ which 
use b-tagging methods similar to those used in the 
OPAL Higgs search~\cite{bib:opal-higgs}.
The use of b-tagging improves the separation
of the $\PZ$-pair signal from background and
allows us to check the $\bbbar$ content of our $\PZ$-pair
sample for consistency with the Standard Model.
The description of the individual selections
is followed by a discussion of possible systematic
errors (Section~\ref{ssec:sys}).
In Section~\ref{sec:xsec} the selected events 
are used to measure the $\PZ$-pair cross section.
Then the cross section and angular distribution
are compared with
the Standard Model predictions and limits
on anomalous neutral current triple gauge couplings
are derived. 

\section{Data analysis and Monte Carlo}           \label{sec:data}

The OPAL detector\footnote{OPAL uses a right-handed coordinate system in
which the $z$ axis is along the electron beam direction and the $x$
axis is horizontal. The polar angle, $\theta$, is measured with respect
to the $z$ axis and the azimuthal angle, $\phi$, with respect to the
$x$ axis.}, trigger and data acquisition system are described fully 
elsewhere~\cite{bib:OPAL-detector,bib:OPAL-SI,bib:OPAL-SW,bib:OPAL-TR,
bib:OPAL-DAQ}. Our analyses use approximately $\lumia~\pb^{-1}$ of data 
collected at center-of-mass energies between 181--184\,GeV 
and approximately 
$\lumib~\pb^{-1}$ collected at center-of-mass energies near
189\,GeV.
The corresponding luminosity-weighted mean center-of-mass energies are
\mbox{$\emna \pm 0.05$\,GeV} and 
\mbox{$\emnb \pm 0.04$\,GeV}~\cite{bib:ELEP}.
The luminosity was measured using small-angle Bhabha scattering events
recorded in the silicon-tungsten 
luminometer~\cite{bib:OPAL-SW,bib:OPAL-SM172,bib:OPAL-SM183}
and the theoretical calculation of Reference~\cite{bib:bhlumi}.
The overall error on the luminosity measurement amounts to 
less than 0.5\% and contributes negligibly to our 
cross-section measurement error. 

Selection efficiencies and backgrounds were calculated 
using Monte Carlo simulations.  
All events were passed through
a simulation~\cite{bib:gopal} of the OPAL detector and processed as for 
data.
We define the $\PZ\PZ$ cross section as the contribution to
the total four-fermion cross section from the NC2
$\PZ$-pair diagrams
shown in Figure~\ref{fig:eetozz}.
All signal efficiencies given in this paper are with
respect to these $\PZ$-pair processes.
Contributions from all other four-fermion final states,
including interference with NC2 diagrams,
are considered as background.
For studies of the signal efficiency
we have used grc4f~\cite{bib:grc4f},
YFSZZ~\cite{bib:yfszz} and 
PYTHIA~\cite{bib:pythia}.

Backgrounds are 
simulated using several different generators.
PYTHIA
is used to simulate
two-fermion final states such as 
$\epem \to \PZ^* (n\gamma) \to \qqbar (n\gamma) $ and 
$\epem \to \gamma^* (n\gamma) \to \qqbar (n\gamma) $,
where $(n\gamma)$ indicates the generation of one or
more initial state
photons. HERWIG~\cite{bib:herwig} and KK2f~\cite{bib:kk2f}
are used as checks for these final states.
These two-fermion generators include gluon radiation
from the quarks which produce $\qqbar {\rm g}$, 
$\qqbar \qqbar$ and
$\qqbar{\rm gg}$ final states.
The grc4f generator, with the contribution 
exclusively due to NC2 diagrams removed,
is used to simulate  other four-fermion background.
KORALW~\cite{bib:koralw} and EXCALIBUR~\cite{bib:excalibur} are
used as checks of the four-fermion background.
%
%
Multiperipheral (``two-photon'') processes with hadronic final
states are simulated by
combining events from PYTHIA, for events without electrons scattered
into the detector, and
HERWIG~\cite{bib:herwig}, for events with electrons scattered
into the detector.
For the $\qqee$ final state,
TWOGEN~\cite{bib:twogen} is used to simulate
two-photon events with both the electron and positron
scattered into the detector.
The  Vermaseren~\cite{bib:vermaseren} generator is
used to simulate
multiperipheral production of the final states $\epem \ellell$.

To avoid background from 
four-fermion final states mediated by
$\PZ\gamma^*$,
our selections were optimized
to select simulated events with masses,
$m_{1}$ and $m_{2}$, that satisfy
\mbox{$ m_{1} + m_{2}     > 170 \ \mathrm{GeV}$} and 
\mbox{$| m_{1} - m_{2} |  <  20 \ \mathrm{GeV}$}.
At 189\,GeV (183\,GeV)
more than 90\% (80\%) of the events
produced via the NC2 diagrams are contained in this mass
region.
Events from the 
NC2 diagrams dominate in this mass region except for
final states containing electron pairs.
Backgrounds in these samples from two-photon
and electroweak Compton scattering
($\Pe \gamma \to \Pe \PZ$) 
processes~\cite{bib:zee} are reduced by using electrons detected in the
electromagnetic calorimeters with 
$| \cos \theta_{\mathrm{e}} |  <  0.985$,
where  $\theta_{\mathrm{e}}$ is the polar angle
of the electron.

\section{Event selection}           \label{sec:sel}

In the following subsections we describe event selections
which exploit every decay mode of the $\PZ$ boson.  Our
selections cover all $\PZ\PZ$ final states except
$\nnnn$ and $\ttnn$. 
In hadronic final states, the energy and direction of the jets
are determined using reconstructed tracks and calorimeter clusters
using the correction for
double counting described in Reference~\cite{bib:opal-higgs}.
In the $\qqll$, $\qqqq$
and $\qqbb$ analyses  
four-constraint (4C) and five-constraint (5C) kinematic fits are used.
The 4C fit imposes energy and momentum conservation.
In the 5C fit the added constraint requires 
the masses  of the two candidate $\PZ$ bosons to be equal
to one another.
For final states with one or more  $\PZ$ bosons decaying
to tau pairs, the energy and total momentum of the
tau leptons are obtained by leaving the reconstructed
direction of the four fermions fixed and scaling
the energy and momentum of each of the fermions
to obtain energy and momentum conservation.  The
scaled values of the tau momentum and energy
are then used in the subsequent steps of the analysis.
In the
$\qqtt$ and $\bbtt$ final states, subsequent kinematic
fits are effectively 2C and 3C fits.


\subsection{Selection of {\mybold $\zztollll$} events }
\label{ssec:llll}
$\PZ$-pair events decaying to final states 
with four charged leptons ($\llll$)
produce low multiplicity events with
a clear topological signature that
is exploited to
maximize the selection efficiency.
The $\llll$ analysis begins by
selecting low multiplicity events (less than
13 tracks or clusters) with visible
energy of at least $0.2 \roots$ and at least
one track with momentum
of 5\,GeV or more. 
Using a cone algorithm,
the events are required to  have 
exactly four cones of $15^\circ$ half angle
containing between 1 and 3 tracks.
Cones of opposite charge are paired\footnote{Two-track 
cones are assigned the charge of 
the most energetic track if
the momentum of one track exceeds that of
the other by a factor of 4.  
Events with
a cone which fails this requirement are rejected.}
to form $\PZ$ boson candidates.

Lepton identification is
only used to classify events as background 
or to reduce the number of cone combinations
considered by preventing the matching of
identified electrons with identified muons.
Electrons are identified on the basis of energy
deposition in the electromagnetic calorimeter,
track curvature and
specific ionization in the tracking chambers.  Muons
are identified using the association
between tracks and 
hits in the hadron calorimeter and muon chambers.

To reduce background from two-photon events with a single
scattered electron detected, we eliminate events
with forward going electrons (backward going positrons)
with the cut \mbox{$ \cos \theta_{\mathrm{e^-}} < 0.85$} 
(\mbox{$ \cos \theta_{\mathrm{e^+}} > -0.85$}).
Here $\theta_{\mathrm{e^-}}$ (  $\theta_{\mathrm{e^+}}$)
is the angle of the electron (positron) with
respect to the incoming electron beam.
Background from
partially reconstructed $\qqbar (n\gamma)$  
events and two-photon events is reduced by
requiring that most of the energy is
not concentrated in a single cone,
$E_{\mathrm{vis}} -  E_{\mathrm{cone}}^\mathrm{max} > 0.2\roots$.
Here $E_{\mathrm{vis}}$ is the total visible energy of the
event and $E_{\mathrm{cone}}^\mathrm{max}$ is the energy
contained in the most energetic cone.

The invariant masses of the lepton pairs are calculated in
three different ways which are motivated by the 
possibility of having zero, one  or two tau pairs 
in the event.
The events are classified according to
the number of tau pairs in the event.
(i) Events with  $\Evis > 0.9 \roots$ 
are treated as
$\eeee$, $\eemm$ or $\mmmm$ events. 
We also treat all events 
with $| \cos \theta_{\mathrm{miss}}|  > 0.98$
($\theta_{\mathrm{miss}}$ is the polar angle
associated with the missing momentum in the
event)
as $\eeee$, $\eemm$ or $\mmmm$ events to
maintain efficiency for $\PZ$-pairs with
initial state radiation.
In these events there are no missing neutrinos and the mass 
of each pair of lepton cones is evaluated.  
(ii) Events failing (i) with a cone-pair combination
that has energy exceeding  $0.9 \mPZ$ are tried
as an $\eett$ or $\mmtt$ final state. 
The mass of the tau-pair system
is calculated from the recoil mass of the presumed
electron or muon pair.
(iii) Any remaining combinations are treated as
$\tttt$ final states. The momenta of the
tau leptons are determined with the scaling procedure
described in the introduction to Section~\ref{sec:sel} 
and the invariant masses of the cone
pairs
are evaluated using the scaled momenta.  
In any event with more than one valid combination, 
each combination is tested
using the invariant mass cuts listed below.

\begin{figure}
  \center{
   \begin{tabular}{cc}
   \epsfig{file=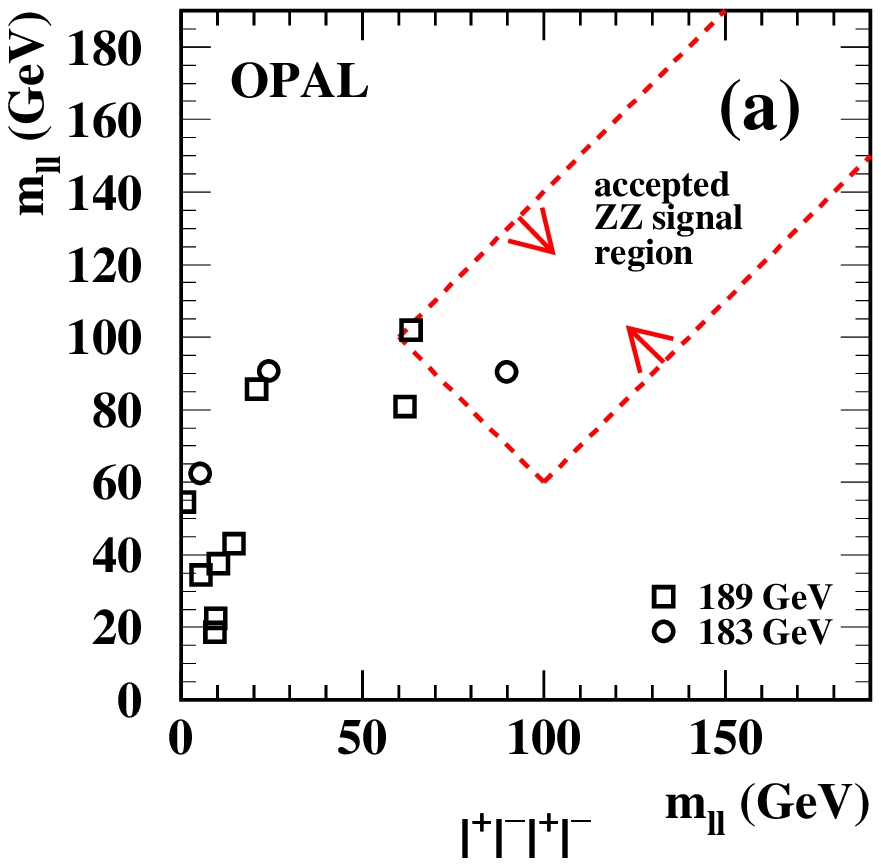,width=0.49\textwidth
      } &
   \epsfig{file=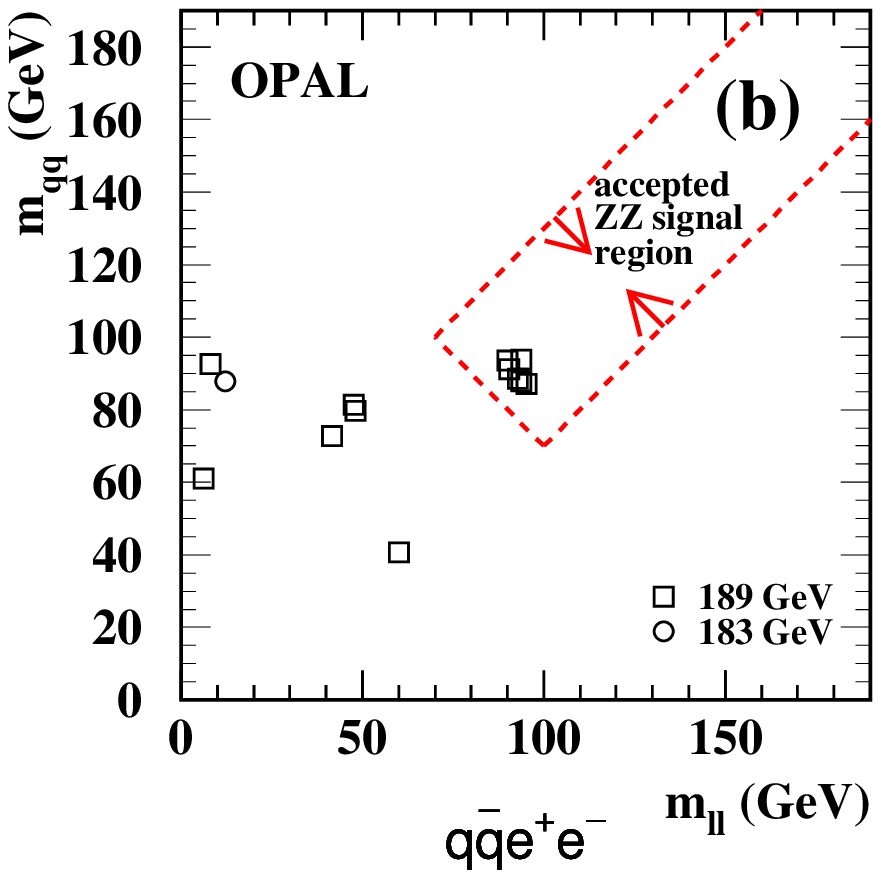,width=0.49\textwidth
      }\\    
    \epsfig{file=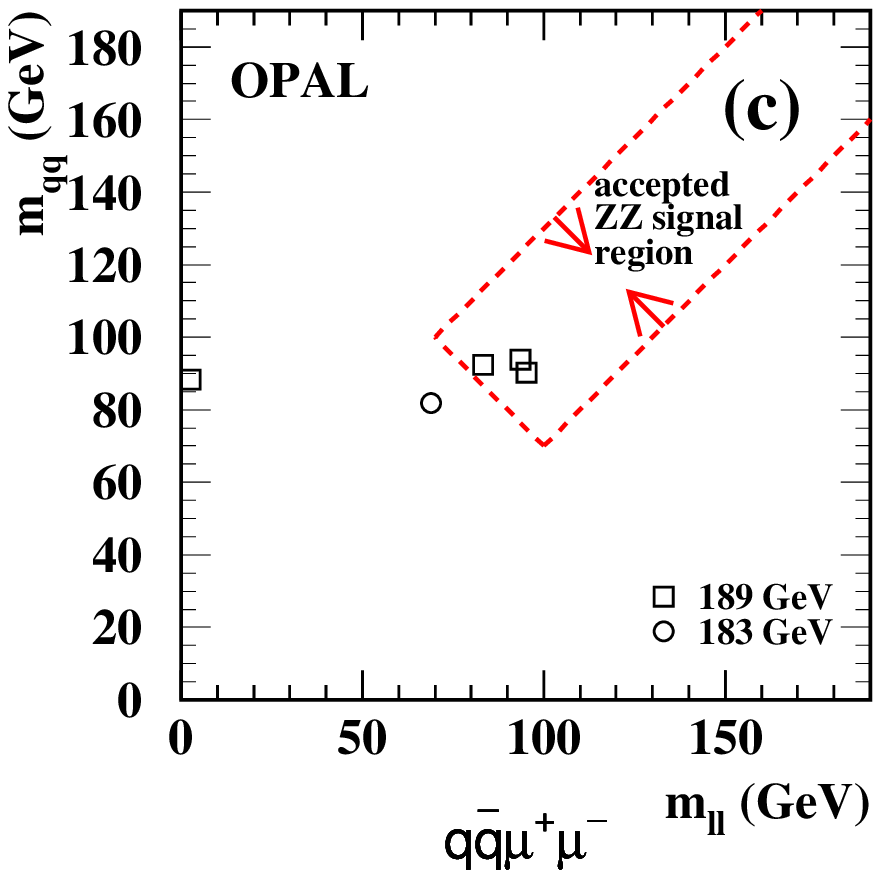,width=0.49\textwidth
      } &
   \epsfig{file=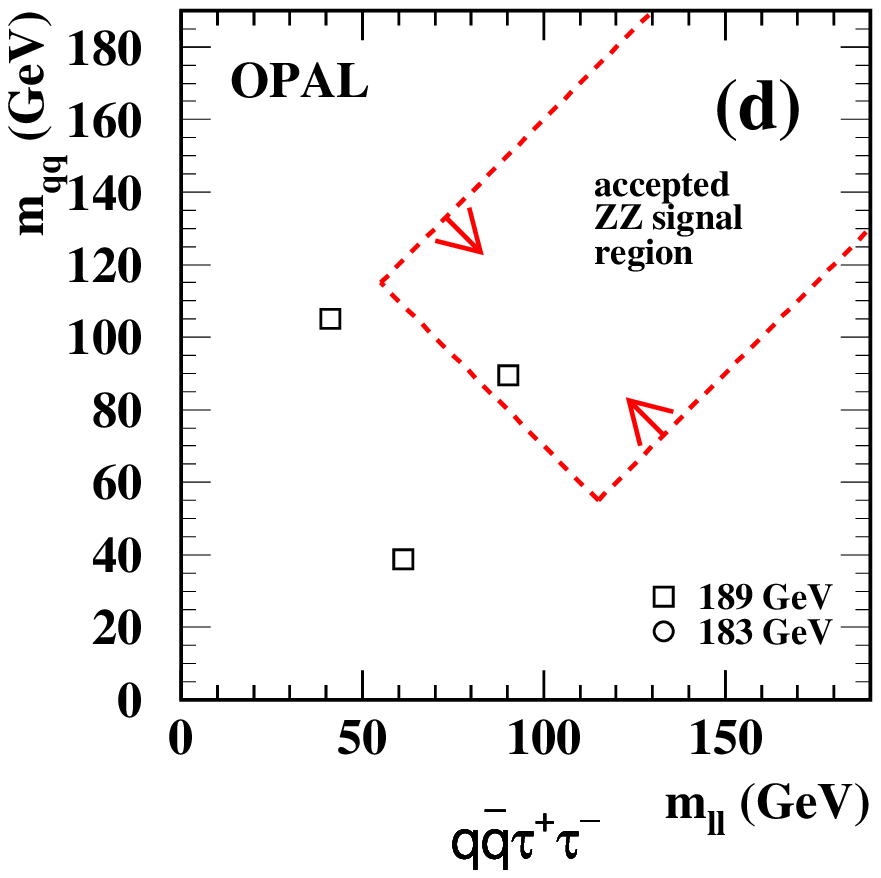,width=0.49\textwidth
      }\\
   \end{tabular}
}
  \caption[ zzllll ]{
  \label{fig:zzmm}
   (a) Invariant masses of $\llll$ cone pairs 
for the data at $\roots = 183$\,GeV and 
 $\roots = 189$\,GeV.
Invariant mass pairs for the 
(b) $\qqee$ 
(c) $\qqmm$  and 
(d) $\qqtt$ data.
The dashed lines show the final invariant mass cuts.
}
\end{figure}

To reduce the combinatorial background, combinations
with pair masses closest to $\mPZ$ are selected.
In events with one or more
combination
satisfying 
$|\mPZ - m_{\ell\ell}| > 0.1 \mPZ$
the cone-pair combination with the smallest
value of $(\mPZ - m_{\ell\ell} )^2 + (\mPZ - m_{\ell'\ell'} )^2$
is selected for further analysis.
In the other combinations, the combination with 
the smallest value of $| \mPZ -  m_{\ell\ell} |$ or
$| \mPZ -  m_{\ell'\ell'} |$ is selected.
The final event sample is then chosen with
the requirement
$m_{\ell\ell} + m_{\ell'\ell'}  >  160$\,GeV and
$| m_{\ell\ell} - m_{\ell'\ell'} |   <  40$\,GeV.
The signal
detection efficiency, averaged over all $\llll$ final
states is given in Table~\ref{tab:zzxsec}~(line~\lllllb).
The efficiency for individual final states range
from 30\% for $\tttt$ to more than 70\% for $\mmmm$.
The invariant masses of all cone pairs passing one of the
selections are shown in 
Figure~\ref{fig:zzmm}a.
One candidate 
is found in the
183\,GeV data and one candidate is found in the 189\,GeV
data.

%
%

\subsection{Selection of {\mybold $\zztollnn$}\ events }
\label{ssec:llnn}
The selection of the $\eenn$ and $\mmnn$ final states
is based on the OPAL selection of
$\PW$ pairs decaying to
leptons~\cite{bib:OPAL-WW}.
The mass and momentum
of the  $\PZ$ boson decaying to $\nunu$
are calculated using the beam
energy constraint and the visible
decay of the other $\PZ$ boson to a charged lepton pair.
A likelihood selection based on the visible and recoil masses
as well as the polar angle of the leptons,
is then used to separate signal from background.
 
The $\eenn$ selection starts with OPAL
$\PW$-pair candidates where both charged
leptons are classified as electrons.
Each event is
then divided into two hemispheres using the thrust axis.
The highest momentum charged (leading) track is selected from each hemisphere.
The sum of the charges of these two tracks is required to be zero.
The determination of the 
visible mass, $\mvis$, and the recoil mass,
$\mrec$, is based on the
energy as measured in the electromagnetic calorimeter and the direction
of the leading tracks.

\begin{figure}
\begin{center}
\epsfig{file=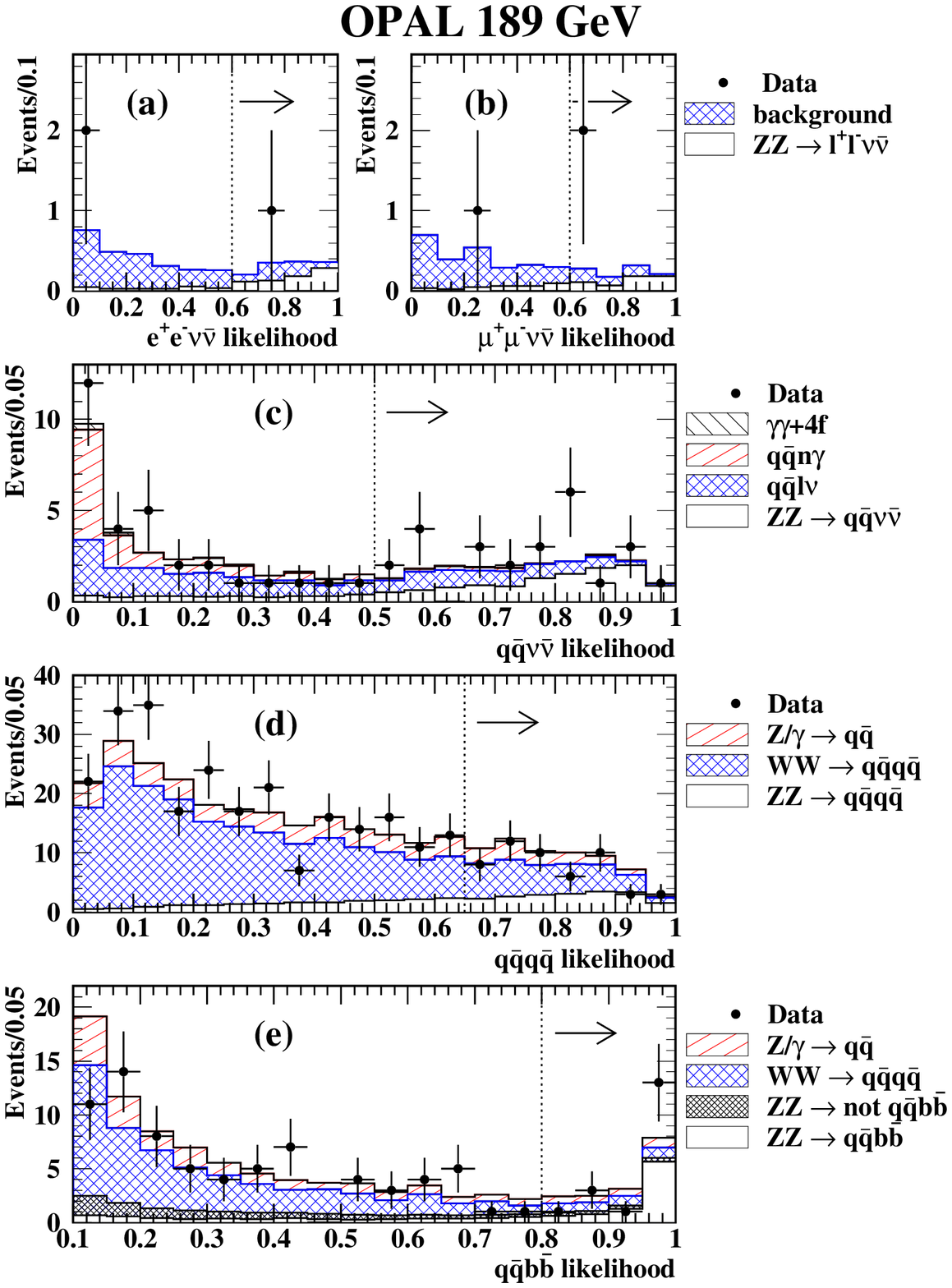,width=0.95\textwidth}
\end{center}
\vspace{-1.0cm}
\caption[$\qqnn$ likelihood ]{
Likelihood discriminant
used at $\roots = 189$\,GeV for 
(a) the $\eenn$ selection,
(b) the $\mmnn$ selection,
(c) the $\qqnn$ selection, 
(d) the $\qqqq$ selection
and (e) the $\qqbb$ selection. 
\label{fig:zzlike}
}
\end{figure}

Three variables were chosen for the likelihood selection: 
$Q\cos\theta$, 
where $\theta$ is the angle of the highest momentum charged track and 
$Q$ is its charge,
the normalized sum of visible and recoil masses $(\mvis+\mrec)/\roots$
and 
the difference of visible and recoil masses, $\mvis-\mrec$.
The performance of the
likelihood is improved with the following preselection:
$\mathrm{-25 \,GeV < \mvis-\mrec < 15 \,GeV}$ and
$\mathrm{(\mvis+\mrec)/\roots >0.90}$.
One event with $\Leenn >0.60$ is selected
(see Table~\ref{tab:zzxsec} (line~\eennlb) and Figure~\ref{fig:zzlike}a).

The $\mmnn$ selection starts 
with the OPAL $\PW$-pair candidates
where both charged leptons are
classified as muons.
The selection procedure is  the same as for the $\eenn$
final states except that $\mvis$, $\mrec$, and $E_{\mathrm{vis}}$
are calculated from the the momentum of the reconstructed tracks
of the $\PZ$ boson decaying to muon pairs.  
The
likelihood preselections
$\mathrm{-25 \,GeV < \mvis-\mrec < 25 \,GeV}$ and
$\mathrm{(\mvis+\mrec)/\roots >0.90}$ are applied.
Two events with $\Lmmnn >0.60$ are selected
(see Table~\ref{tab:zzxsec} (line~\mmnnlb) and Figure~\ref{fig:zzlike}b).

\subsection{Selection of {\mybold $\zztoqqll$}\ events}
\label{ssec:qqll}
The 
lepton pairs in the $\qqbar\epem$\ 
and $\qqbar\mumu$\ final states
have a distinctive signature 
making possible selections with high 
efficiencies and a low background contamination.  
In the $\qqtt$ final state,
the decay of the tau leptons produces
events which are more difficult to identify.
The identification of this final state exploits
the missing momentum and
missing energy carried away by the neutrinos
produced in the decay of the tau lepton.

\subsubsection{Selection of {\mybold $\zztoqqee$}\ 
and  {\mybold $\zztoqqmm$} events}

The selection of $\qqee$ and $\qqmm$ final states 
requires the visible
energy of the events
to be greater than 90\,\mbox{GeV} and at least six 
reconstructed tracks. 
Among all tracks with
momenta greater than 2\,GeV, the highest momentum track is taken as
the first lepton candidate and the second-highest momentum track
with a charge opposite to the first candidate is taken as the second
lepton candidate. 
Using the Durham~\cite{bib:durham} jet
algorithm, the event, including the lepton candidates, 
is forced into four jets  
and the jet  resolution variable that separates the three-jet
topology from the four-jet topology,
$y_{34}$, is required to be greater than $10^{-3}$. 
Excluding the electron or muon candidates
and their associated calorimeter clusters, the rest of the event is
forced into two jets.  The 4C and 5C fits to
the two lepton candidates and the two jets are
required to converge.\footnote{ In the context of
this paper, convergence is defined as a
fit probability greater than $10^{-10}$.}

In the $\qqee$ selection 
no explicit electron identification
is used. Electron candidates are selected by 
requiring the sum of the electromagnetic cluster
energies $E_1 + E_2$ associated to the electrons to be greater than
70\,GeV and the momentum of the most energetic electron track
to exceed 20\,GeV.
We also reject the event if
the angle between either electron candidate 
and any other track is less than 5$^\circ$.  

In the $\qqmm$ selection the muons are identified using
(i) tracks which match a reconstructed segment in
the muon chambers,
(ii) tracks which are associated to hits in the hadron calorimeter
or muon chambers~\cite{bib:OPAL-WW},
or 
(iii) isolated tracks associated to
electromagnetic clusters with reconstructed energy less than 2\,GeV.
No isolation requirement is
imposed on events with both muon tracks passing (i) or (ii).
Events with at least one muon identified with criterion (iii)
are accepted if both muon candidates in the
event have an angle of at least 10$^\circ$ to the 
nearest track.
We require the sum of the momenta of the
two leptons to be greater than 70\,GeV.

$\PZ$-pair events are separated from $\PZ\gamma^*$ background
by requiring
the fitted mass of the 5C fit
to be larger than 85\,GeV
and the invariant masses $\mll$ and $\mqq$ obtained
from the 4C fit to satisfy
 $ (\mll + \mqq)    >  170$\,GeV and
 $|\mll - \mqq |  <   30$\,GeV.
Figure~\ref{fig:zzmm}b (\ref{fig:zzmm}c) shows the
distribution of $\mee$\ ($\mmumu$) and $\mqq$\ before the cuts on the
masses from the 4C and 5C fits.

After all cuts the selection efficiency\footnote{
Small amounts of feedthrough from other $\PZ\PZ$ final states, in
this case $\qqtt$, are counted as signal.} 
for
$\qqee$
signal events is $\meqqeea$\% at 183\,GeV and
$\meqqeeb$\% at 189\,GeV.
The errors on these efficiencies include the systematic
errors (see Section~\ref{ssec:sys}). 
No candidate is observed at 183\,GeV.
Six candidate events are found after all cuts in the data taken at 189\,GeV.  
In Table~\ref{tab:zzxsec} (lines~\qqeelb\ and \bbeelb)
we give the efficiency, background
and observed number of events.
The largest source of background after all
cuts is from $\PZ\gamma^*$ mediated $\qqee$\ events and from two-photon
events.

For the $\qqmm$ final state the 
selection efficiency is $\meqqmma$\% at
183\,GeV and $\meqqmmb$\% at 189~GeV.  Three events are observed in
the 189~GeV data sample and none in the 183\,GeV data sample.
The errors on these efficiencies include the systematic
errors given below. 
In Table~\ref{tab:zzxsec} (lines~\qqmmlb\ and \bbmmlb)
we give the efficiency, backgrounds
and observed number of  events.
The 
background after all cuts is expected to come
from $\eetozg\to\qqmm$\ events.

\subsubsection{Selection of {\mybold $\eetozz\to\qqbar\tautau$}\ events}

The $\qqtt$ final state is selected from a
sample of events with track
multiplicity greater or equal to six.  
Events which have been
selected as $\eetozz\to\qqbar\epem$ or $\eetozz\to\qqbar\mumu$ are
excluded from this selection.
The tau-lepton
candidates are selected using an artificial neural network algorithm
which is described in detail in Reference~\cite{bib:opal-higgs}.  
The tau candidate with 
the highest neural network output value is taken as the
first candidate. The second best candidate is required to have its
charge opposite to the first candidate and the highest output
value among all remaining candidates.  If a second candidate cannot be
found the event is rejected.  

Motivated by the presence of neutrinos in the final state,
the visible energy of the event, $E_{\mathrm{vis}}$,  
is required to exceed 90\,GeV
and the missing energy $\roots - E_{\mathrm{vis}}$ is
required to exceed 15\,GeV.
In addition, the sum of the momenta of the leading
tracks from the tau-lepton 
decays is required to be less than 70\,GeV. 
Since the direction of the
missing momentum in signal events will tend to be along the direction
of one of the decaying tau leptons, the angle
$\alpha_{\tau,{\rm miss}}$ between the missing momentum and a 
tau-lepton
candidate is required to satisfy $\alpha_{\tau,{\rm miss}} < 90^\circ$ 
for at
least one of the two tau candidates. 

The two hadronic jets are selected in
the same way as in the $\eetozz\to\qqbar\epem$ selection.
The 
initial estimate of the energy and the momenta of the tau
candidates is found from the sum of the
tracks associated to the tau by the
neural network algorithm and all unassociated electromagnetic clusters
in a cone with a half angle of 10$^\circ$ around the leading 
track from the tau decay.  
A 2C kinematic fit that imposes energy and momentum
conservation (see the introduction to
Section~\ref{sec:sel}) is required to converge.
A 3C kinematic fit, with the additional constraint 
of the equality of the fermion pair masses is also  required to converge.

Using the network output~\cite{bib:opal-higgs} for each
tau lepton, a probability is calculated taking into account the different
branching ratios, sensitivities, efficiencies and background levels
for 1-prong and 3-prong tau-lepton decays.
In the following, we
combine the probabilities ${\cal P}_1$ and ${\cal P}_2$
to form a likelihood using
\begin{equation}
{\cal L} = \frac{{\cal P}_{1}{\cal P}_{2}}
{{\cal P}_{1}{\cal P}_{2} + (1-{\cal P}_{1})(1-{\cal P}_{2})}  .
\label{eqn:ttprob}
\end{equation}
The likelihood associated with probabilities of the
two tau candidates is required to 
satisfy ${\cal L}_{\tau\tau} > 0.977$.
In addition, the common mass of the 3C fit is required to
exceed 85 GeV.
Using the 2C fit masses of the tau pair, $\mtautau$, and the 
quark pair, $\mqq$, as obtained from the kinematic fit, we also require 
$\mqq + \mtautau   > 170$\,GeV and
$| \mqq - \mtautau | < 60$\,GeV.

After all cuts the selection efficiency for signal events is found to
be $\meqqtta$\% at 183\,GeV and $\meqqttb$\% at 189\,GeV. One
candidate event is found in the 
data at 189\,GeV while no candidate is selected at 183\,GeV. 
Figure~\ref{fig:zzmm}d shows the masses of the candidate
events before the invariant mass cuts.
In Table~\ref{tab:zzxsec} (lines~\qqttlb\ and \bbttlb)
we give the efficiencies, backgrounds
and observed number of events.

\subsubsection{Selection of {\mybold $\eetozz\to \bbll$} }

Events with $\bbbar$ final states are selected using the algorithm
described in Reference~\cite{bib:opal-higgs}.  
The b probabilities of the two hadronic jets
are combined to form a
likelihood, ${\cal L}_{\mathrm{bb}}$,  
according to Equation~\ref{eqn:ttprob}.
Because the 
$\qqee$ and 
$\qqmm$ selections are pure, a relatively loose cut
of $ {\cal L}_{\mathrm{bb}} > 0.2$ is used to
select the $\bbee$ and $\bbmm$ samples.
For the selections with electron and muon pairs 
there are two classes of events since the selected
$\bbll$ events are a subset of the 
$\qqll$ events.
In Table~\ref{tab:zzxsec} (lines~\bbeelb\ and \bbmmlb)
we give the efficiencies of the
b-tagged samples with respect to the expected fraction of 
$\bbbar$ events.  The efficiencies for samples without b-tags
are given with respect to the hadronic decays
without $\bbbar$ final states.

In the $\bbee$\ selection one candidate is found in the data.
The selection efficiency
is found to be (47$\pm$3)\% at 183\,GeV and
(46$\pm$3)\% at 189\,GeV.
\footnote{
The efficiencies given in this section do not
include feedthrough from other $\qqll$ final
states.  The efficiencies given in 
Table~\ref{tab:zzxsec}
include this feedthrough.}
In the $\bbmm$\ selection we find no candidate at 183\,GeV and one
candidate at 189\,GeV. 
The selection efficiency is found to
be (50$\pm$3)\% at 183\,GeV and (54$\pm$3)\% at 189\,GeV.

For the $\bbtt$ selection the ${\cal L}_{\mathrm{\tau\tau}}$ 
cut of the $\qqtt$ selection is
loosened and combined with ${\cal L}_{\mathrm{bb}}$ as follows.
${\cal L}_{\tau\tau}$ and ${\cal L}_{\mathrm{bb}}$ are both
required to be greater than 0.1.
The  $\bbbar\tautau$ probability for the event,
${\cal L}_{\mathrm{bb} \tau\tau}$, is calculated from
Equation~\ref{eqn:ttprob} with
${\cal L}_{\tau\tau}$ and  ${\cal L}_{\mathrm{bb}}$ as inputs
and required to
exceed 0.95. 
After the cut on ${\cal L}_{\mathrm{bb} \tau\tau}$,
the remaining cuts of the $\qqtt$\ selection are applied,
giving a selection efficiency of $(21 \pm 3)$\% at 183\,GeV
and $(24 \pm 3)$\% at 189\,GeV.
No candidate event is found in the 183\,GeV or 189\,GeV data.

We also use an alternative 
jet-based
$\bbtt$ selection~\cite{bib:opal-higgs} and accept any event which passes
either $\bbtt$ analysis, but 
events previously selected by another $\qqll$ selection
are rejected.
The alternative analysis uses a different approach to reconstruct the 
tau leptons. 
This event selection consists of a set of preselection cuts and 
a subsequent multivariate likelihood selection. 
Events are reconstructed as four jets 
using the Durham algorithm.
Tau-lepton candidates are sought in the four jets
using a likelihood technique to separate real tau leptons and fakes 
in quark jets. The tau and b-tag likelihood values are combined in
a $\bbtt$ likelihood which is maximized to choose the b jets and
tau leptons of the event. 
This $\bbtt$ likelihood
uses tau  and b-tag likelihoods and some topological variables as
input.
Events are accepted if their likelihood exceeds 0.6.
In addition, the fitted 2C masses are required to satisfy
$\mqq + \mtautau  > 170$\,GeV and 
$| \mqq - \mtautau | < 60$\,GeV.

At $189\,\mathrm{GeV}$ data 
the alternative selection has an efficiency of 
$(30\pm 3)$\% for $\bbtt$ events.  
After combining the two selections, the efficiency for
$\bbtt$ events for all cuts except those rejecting
events found by the other $\qqll$ selections is 40\%.  
A similar improvement is realized for the $183\,\mathrm{GeV}$ data.
The efficiency and backgrounds after rejecting
events found by the other selections are given
in Table~\ref{tab:zzxsec} (line~\xbttlb).
One exclusive event is selected by the combined $\bbtt$ analysis
at $189\,\mathrm{GeV}$.

\subsection{Selection of {\mybold $\zztoqqnn$}\ events }
\label{ssec:qqnn}
The $\qqnn$ selection is based on
the reconstruction of the $\PZ$ boson
decaying to $\qqbar$ which produces somewhat back-to-back jets.  
The
selection uses contained events
with a two-jet topology.
The beam
energy constraint is then used to determine the mass
of the  $\PZ$ boson decaying to $\nunu$.  
The properties of the $\qqbar$ decay and the inferred mass
of the $\nunu$ decay are then used in a likelihood analysis
to separate signal from background.

Two-jet events are selected by
dividing each event into two hemispheres using the 
plane perpendicular to the thrust
axis.
The number of charged tracks in each hemisphere is required
to be four or more.
The
polar angles  of the energy-momentum vector 
associated with each hemisphere, 
$\theta_{\mathrm{hemi1}}$
and $\theta_{\mathrm{hemi2}}$, are used to calculate 
the quantity 
${\cos\theta_{\mathrm h}} =  
       \frac{1}{2}(\cos\theta_{\mathrm{hemi1}}
- \cos\theta_{\mathrm{hemi2}})$. 
Contained events are selected by requiring
$|{\cos\theta_{\mathrm{h}}}| < 0.80$. 
The total energy in the forward detectors and
in the forward region of the electromagnetic
calorimeter ($|\cos\theta|>0.95$) is required to be less than 3\,GeV.
$\PW$ decays identified  by the OPAL $\PW$-pair selection are rejected;
the likelihood for $\epem \to \qqlnu$ from Reference~\cite{bib:OPAL-WW},
$\LWW$, is required to be 0.5 or less.

An important background to our selection is
$\qqbar (n\gamma)$ events with photons which escape 
detection.  We discriminate against these events
by looking for a significant amount of
missing transverse momentum, $\pt$.
In each event, $\pt$ can be resolved into
two components, $\pti$, perpendicular to both 
the thrust axis and the beam axis and 
$\ptj$, along the thrust axis and perpendicular
to the beam axis.  
Since $\pti$ is based primarily on angular
measurements, it is better measured than $\ptj$.  
We approximate $\pti$ as 
$\pti  = \frac{1}{2} \Ebeam \sin\phi \sin\theta_{\mathrm h}$.
Here $\Ebeam = \roots /2$ is the beam energy,
$\phi$ is the acoplanarity of the momentum
vectors of the two hemispheres
and  
$\sin\theta_{\mathrm h} = 
\sqrt{ 1 - \cos^2 \theta_{\mathrm h} }$.
The resolution on $\pti$, $\sigma_{\pti}$, was parameterized
as a function of  thrust and ${\cos \theta_{\mathrm h}}$ using
data taken at the $\PZ$ resonance.
The variable
$R_{\pti} = (\pti - \pti^0)/\sigma_{\pti}$
is used as an input to the likelihood.
Here $\pti^0$ corresponds to the transverse
momentum carried by a
photon with half the beam energy 
which just misses the inner edge of our acceptance 
( $\pti^0 = \Ebeam \sin(32\,\mbox{mrad})/2$).
We also use the variable 
$\cos\theta_{\mathrm{miss}}$, the direction of the
missing momentum in the event, to discriminate
against the $\qqbar (n\gamma)$ events.

In the final selection of events, we use a likelihood based
on the following five variables:
(i) the normalized 
sum of visible and recoil masses $(\mvis + \mrec)/ \roots $,
(ii) the difference of visible and recoil masses ($\mvis - \mrec$),
(iii)
$\log(y_{23})$, where $y_{23}$ is the jet resolution parameter 
that separates the two-jet
topology from the three-jet topology as
calculated from the Durham jet algorithm,
(iv)
$\cos\theta_{\mathrm{miss}}$ and
(v) $R_{\pti}$.
The mass variables are useful for reducing background from
$\PW$-pair production and $\Wenu$ final states. 
The jet resolution parameter is useful in reducing
the remaining $\qqlnu$ final states.
To improve the performance of the likelihood analysis we
use only events with : $| \mvis - \mrec | < 50$\,GeV,
$(\mvis + \mrec)/ \roots > 0.89$ and $R_{\pti}>1.2$.
Events are then selected using
$\Lqqnn > 0.5$, where $\Lqqnn$ is the likelihood for
the $\qqnn$ selection.
The likelihood distribution of data and Monte Carlo is shown in
Figure~\ref{fig:zzlike}c. 
For the $\bbnn$ selection we require, in addition,
the b-tag variable of 
Reference~\cite{bib:opal-higgs} to be greater than 0.65.

The efficiencies for the $\qqnn$ selection alone are
$\meqqnna$\% at 183\,GeV and 
$\meqqnnb$\% at 189\,GeV.
The errors on these efficiencies include the systematic
errors discussed below in Section~\ref{ssec:sys}.  
The efficiencies after considering the results of
the b-tagging, as well as the
number of events selected at the two energies are given in
Table~\ref{tab:zzxsec} (lines~\qqnnlb\ and \bbnnlb).

\subsection{Selection of {\mybold $\zztoqqqq$}\ events }
\label{ssec:qqqq}
The fully hadronic channel of the $\PZ$-pair decay has the largest
branching fraction of all channels (about 50\%), but suffers from large
background from hadronic $\PW$-pair decays.
We apply two different event selections, one of which is based
mainly on reconstructed mass information in order to
accept all hadronic $\PZ$-pair decays without flavor requirement, while a
second analysis applies a flavor tag in order to select
final states involving b quarks, allowing for looser requirements
on the reconstructed boson mass.

For both subsamples, hadronic $\PZ^*/\gamma^* \to \qqbar$ events are an
important background. We therefore start with a common preselection
based on event shape variables which is mainly aimed at reducing this
background.
We use a likelihood method, described below
in Section~\ref{sec:jetpair}, in order to choose the most likely
jet pairing for $\PZ$-pair decays.

\subsubsection{Preselection and jet pairing}
\label{sec:jetpair}
The event selection starts from the inclusive multihadron selection
described in Reference~\cite{bib:OPAL-SM172}\@.
      The radiative process $\epem \to \PZ \gamma \to \qqbar \gamma$ 
      is suppressed
      by requiring the effective center-of-mass energy after
      initial state radiation,
      $\rootsp$, to be larger than 150\,GeV\@.
      $\rootsp$ is  obtained from
      a kinematic fit~\cite{bib:OPAL-SM172} that allows 
      for one or more radiative
      photons in the detector or along the beam pipe.
      The final state particles are then grouped into jets using
      the Durham algorithm~\cite{bib:durham}.
      A four-jet sample is formed by requiring 
      the jet resolution parameter $y_{34}$ to be at least
      $0.003$ and each jet to contain at least two charged
      tracks.
      In order to suppress $\PZ^*/\gamma^* \to \qqbar$ background,
      the event shape parameter $C_{\mathrm{par}}$
      \cite{bib:cpar}, which
      is large for spherical events, is required to be greater than 0.25.
      A 4C kinematic
      fit using energy and momentum conservation
      is required to converge. 
      A 5C kinematic fit which forces the two jet pairs to have the same mass
      is applied in turn to all three possible
      combinations of the four jets.  This fit is
      required to converge for at least one combination.
The efficiencies of these preselection cuts are (86.4 $\pm$ 0.5) \% 
and (88.9 $\pm$ 0.5) \% for signal events at 183\,GeV and 189\,GeV,
respectively.


In order to determine which pair of jets comes from each $\PZ$,
we calculate a likelihood function using
the mass obtained from the 5C fit, the corresponding fit probability,
and the difference between the two di-jet masses obtained from the 4C fit.
In the YFSZZ simulation of $\PZ$-pair decays
the correct jet pairing has the highest likelihood output in 
(86.8 $\pm$ 0.5) \%
of the events.
This fraction rises to 
(93.8 $\pm$ 0.5) \% for the 
events after the final selection.


\subsubsection{Likelihood for the inclusive 
{\boldmath $\PZ\PZ \to \qqbar\qqbar$} event selection}
We use a likelihood selection with eight input variables for the
selection of
$\PZ\PZ \to \qqbar\qqbar$ events.
The first variable is the jet pairing likelihood described above.
Excluding the jet pairing with the largest difference between
 the two di-jet masses as obtained from the 4C fit, we identify
 among the remaining two possible pairings the one
 for which the 5C-fit mass is closer to the $\PW$ mass.
 We use the difference between this 5C-fit mass and the
 $\PW$ mass in order to discriminate against hadronic  
 $\PW$-pair events.
Two variables that are sensitive to unobserved particles along the
beam direction are the fitted center-of-mass energy and the
sum of the cosines of the polar angles of the four jets.
In order to discriminate against $\PZ^*/\gamma^*$ events,
we use the difference between the largest and
smallest jet energies after the 4C fit, and the angular variable
$j_{\rm ang}=E_4(1-\cos\theta_{12}\cos\theta_{13}\cos\theta_{23})/\roots$,
where $E_4$ is the smallest of the four jet energies, and the $\theta_{ij}$
are the opening angles between jets $i$ and $j$, with the jets
ordered by energy.
Finally we calculate from the momenta configuration of the four jets
 the effective matrix element for the QCD processes
$\PZ^*/\gamma^\star\to\qqbar{\rm gg}$ and 
$\PZ^*/\gamma^\star\to\qqbar\qqbar$
as defined in Reference~\cite{bib:seymour},
and
the matrix element for the process
${\rm WW}\to\qqbar\qqbar$ from Reference~\cite{bib:excalibur}\@.

The distribution of the likelihood function calculated from these eight
variables is shown in Figure~\ref{fig:zzlike}d\@.
In order to maximize the significance of the measured cross section,
assuming a 10\% relative systematic error on the background, we place
a cut on the likelihood at 0.65\@. 
This cut leads to an efficiency of
$\meqqqqb$\% (189\,GeV) 
relative to all fully hadronic NC2 $\PZ$-pair final states.
In the data, 52 events are selected.
We expect a total of 57.0 events from Standard Model processes,
of which 17.6 originate from the $\PZ$-pair signal, while 27.0 events
are expected from hadronic $\PW$-pair decays and 12.4 events from
hadronic two-fermion processes.
At 183\,GeV the selection
efficiency for fully hadronic $\PZ$-pair decays
is $\meqqqqa$\%. 
In the data 8 events are selected.
The Standard Model expectation is
1.7 signal events and 7.2 background
events.

\subsubsection{Likelihood for 
{\boldmath $\PZ\PZ \to \qqbar\bbbar$} event selection}

Jets originating from b-quarks are selected using the same
b-tagging algorithm used in the $\bbll$ and $\bbnn$ selections.
We evaluate the probability for each of the four jets to originate from
a primary b quark, and use the two highest probabilities as input variables
for a likelihood to select $\PZ\PZ \to \qqbar\bbbar$ events.
In addition,
we use the parameters $y_{34}$, $C_{\mathrm{par}}$, the difference
between the largest and smallest jet energies and the output of the jet
pairing likelihood. We also use the fit probabilities of a 5C kinematic
fit which constrains one boson mass to the $\PZ$ mass, and the probability
of a 6C fit which forces both masses to be equal to the $\PW$ mass.

Figure~\ref{fig:zzlike}e shows
the distribution of the likelihood function calculated from these
eight variables for the preselected events.
The signal likelihood is required to be larger than 0.80
for both 183\,GeV and 189\,GeV data.
After the likelihood selection, we perform
an additional cut on the mass obtained from the 5C-mass fit 
in the most likely jet pairing, which is required to be larger than 86\,GeV\@.
This choice of the cuts on the likelihood and on the 5C-fit mass
was made by maximizing
the expected statistical significance of signal over
background.
The final efficiency is
$\meqqbba$\% for the 183 GeV data and
$\meqqbbb$\% for the 189 GeV data.
The observed number of events,
expected signal and background are given
in Table~\ref{tab:zzxsec} (line~\qqbblb).

\subsubsection{Combination}

To account for overlap between the $\qqqq$ and $\qqbb$ selections
we divide the data into three logical classes, 
exclusive $\qqqq$ events (Table~\ref{tab:zzxsec} (line~\qqqqlb)),
exclusive $\qqbb$ (line~\qqxblb) events and 
the overlapping region (Table~\ref{tab:zzxsec} (line~\qqbblb)).  
In Table~\ref{tab:zzxsec} (line~\qqqqlb) 
we give the efficiency relative to all fully hadronic final states
without b-quarks.  The other efficiencies are
given relative to fully hadronic final states with b-quarks.

\subsection{Selection systematic errors}
\label{ssec:sys}

Systematic errors have only a modest effect on our final 
result because of the large statistical
error associated with the small $\PZ$-pair cross section.

Detector effects can best be studied by comparing calibration
data taken at the $\PZ$ resonance with a simulation of
the same process.  These comparisons are important
for final 
states such as  $\llnn$, $\qqnn$, 
and $\qqqq$, 
where the tight
cuts are needed to separate signal and background.
In these
cases, we add additional smearing to the energy and momentum
of the simulated events to match data and simulation.
We then apply the same smearing to the signal and background
Monte Carlos and then correct our efficiencies and background
accordingly.  
The full difference is used as the systematic
error in these cases.  
At $\roots = 189$~GeV,
these differences give relative systematic
errors on the efficiency of 2.5\%
for the $\eenn$
final state, 5.2\%  for the $\mmnn$ final state and
3.8\% for $\qqnn$ final state.
In the $\qqqq$ inclusive analysis, where uncertainties
on the reconstructed angles are also important, similar studies 
lead to a relative detector systematic error of 6\%.

Detector systematic 
errors for the $\qqll$ and $\llll$ selections 
without $\tau$-pairs
in the final state are small because of the good separation
of signal and background.  
In the $\llll$ final state the
largest effect (3\%) is
from modeling of the multiplicity requirement which 
is important for final states containing $\tau$-pairs.
In the $\qqtt$ selection, the systematic
uncertainties in the  efficiencies were determined by
overlaying hadronic and tau decays taken from $\PZ$
resonance data giving a contribution to the 
systematic error of 6.0\%.
When the final states are combined to 
determine the cross section and limits on the
anomalous triple gauge couplings, we assume a 
common relative systematic error of 3\% on the efficiencies.

Another important detector effect comes from the simulation
of the variables used by the OPAL b-tag which is discussed in
Reference~\cite{bib:opal-higgs}.  We allow for a common
5\% error in the efficiency of the b-tag, consistent with our
studies on $\PZ$ resonance data and Monte Carlo.

These detector effects were propagated through
to our background errors. 

In each channel the signal and background Monte Carlo 
generators have been compared against alternative
generators.  
In almost all cases
the observed differences are consistent within the finite
Monte Carlo statistics and the systematic error has been
assigned accordingly.
One notable exception is
in the $\qqbb$ background 
where differences in 
the PYTHIA, grc4f and EXCALIBUR simulation of the
$\PW$-pair background are as large as 20\%.  In this
case the full difference has been assigned as the systematic error.

In the $\llll$ channels, which has
a large background from two-photon events,
we have compared the number of selected events
at an early stage of the analysis with the Monte Carlo prediction
and based our background systematic error on the level of
agreement.  This results in 20\% systematic error.

\section{Results}                    \label{sec:xsec}

We combine the information from all of the analyses
reported above using a maximum likelihood fit
to determine the production cross section for
$\eetozz$.
The information which was used in the fit, as
well as the Standard Model prediction for $\PZ$-pair
production, 
is summarized in Table~\ref{tab:zzxsec}.  
For each channel the table gives the number of events observed, 
$\nobs$, the Standard Model prediction for
all events, $\smtot$, the expected
signal, $\nsm$, 
the expected background, $\nback$, 
the efficiency $\eff$, 
and the integrated luminosity, $\lint$.
$\br$ is the branching ratio of $\PZ$-pairs to 
the given final state, calculated from $\PZ$
resonance data~\cite{bib:pdg}.
In the table we give the overlap between 
the b-tag and non-b-tag analyses. 
Possible overlap between $\qqqq$ and $\qqll$ has been studied,
and found to be
an order of magnitude smaller than the overlap of 
$\qqqq$ and $\qqbb$ and has therefore been ignored.

\begin{table}
\begin{center}
\vskip 0.2cm
$\roots = 183$\,GeV \\
\vskip 0.2cm
\begin{tabular}{|c|l|ccccccc|}
\hline
&Selection & $\nobs$ & $\smtot$ &$\nsm$ & $\nback$ &$\eff$ & $\br$ & $\lint$\\
&          &         &          &       &          &       &       &
 $(\pb^{-1})$  \\
\hline
 a&$\llllz$&  1&$ 0.15 \pm  0.03$ & $ 0.07 \pm  0.01$ & $ 0.08 \pm  0.03$ & $0.54 \pm 0.02$ &  0.010&  56.7 \\ 
 b&$\eennz$&  0&$ 0.15 \pm  0.04$ & $ 0.06 \pm  0.01$ & $ 0.09 \pm  0.03$ & $0.34 \pm 0.05$ &  0.013&  56.8 \\ 
 c&$\mmnnz$&  0&$ 0.06 \pm  0.02$ & $ 0.04 \pm  0.01$ & $ 0.03 \pm  0.02$ & $0.20 \pm 0.03$ &  0.013&  56.8 \\ 
 d&$\qqeez$&  0&$ 0.41 \pm  0.04$ & $ 0.27 \pm  0.01$ & $ 0.14 \pm  0.04$ & $0.56 \pm 0.03$ &  0.037&  54.7 \\ 
 e&$\bbeez$&  0&$ 0.10 \pm  0.02$ & $ 0.07 \pm  0.01$ & $ 0.03 \pm  0.02$ & $0.52 \pm 0.05$ &  0.010&  54.7 \\ 
 f&$\qqmmz$&  0&$ 0.40 \pm  0.03$ & $ 0.34 \pm  0.01$ & $ 0.07 \pm  0.03$ & $0.70 \pm 0.02$ &  0.037&  54.7 \\ 
 g&$\bbmmz$&  0&$ 0.08 \pm  0.01$ & $ 0.08 \pm  0.01$ & $ 0.01 \pm  0.01$ & $0.57 \pm 0.05$ &  0.010&  54.7 \\ 
 h&$\qqttz$&  0&$ 0.20 \pm  0.03$ & $ 0.11 \pm  0.01$ & $ 0.09 \pm  0.03$ & $0.23 \pm 0.02$ &  0.037&  54.7 \\ 
 i&$\bbttz$&  0&$ 0.03 \pm  0.01$ & $ 0.03 \pm  0.01$ & $ 0.00 \pm  0.01$ & $0.19 \pm 0.03$ &  0.010&  54.7 \\ 
 j&$\xbttz$&  0&$ 0.02 \pm  0.01$ & $ 0.02 \pm  0.01$ & $ 0.00 \pm  0.01$ & $0.13 \pm 0.03$ &  0.010&  54.7 \\ 
 k&$\qqnnz$&  2&$ 2.62 \pm  0.25$ & $ 0.87 \pm  0.07$ & $ 1.75 \pm  0.24$ & $0.30 \pm 0.02$ &  0.219&  54.5 \\ 
 l&$\bbnnz$&  1&$ 0.42 \pm  0.06$ & $ 0.25 \pm  0.02$ & $ 0.17 \pm  0.06$ & $0.31 \pm 0.03$ &  0.061&  54.5 \\ 
 m&$\qqqqz$&  8&$ 7.43 \pm  0.70$ & $ 1.03 \pm  0.07$ & $ 6.39 \pm  0.69$ & $0.26 \pm 0.02$ &  0.300&  54.7 \\ 
 n&$\qqxbz$&  3&$ 1.88 \pm  0.31$ & $ 0.47 \pm  0.02$ & $ 1.42 \pm  0.31$ & $0.15 \pm 0.01$ &  0.235&  54.7 \\ 
 o&$\qqbbz$&  0&$ 1.41 \pm  0.19$ & $ 0.61 \pm  0.03$ & $ 0.80 \pm  0.19$ & $0.20 \pm 0.01$ &  0.235&  54.7 \\ 
\hline
\end{tabular}
\vskip 0.4cm
$\roots = 189$\,GeV \\
\vskip 0.2cm
\begin{tabular}{|c|l|ccccccc|}
\hline
&Selection & $\nobs$ & $\smtot$ &$\nsm$ & $\nback$ &$\eff$ & $\br$ & $\lint$\\
&          &         &          &       &          &       &       &
 $(\pb^{-1})$  \\
\hline  
 a&$\llllz$&  1&$ 1.02 \pm  0.21$ & $ 0.68 \pm  0.02$ & $ 0.35 \pm  0.21$ & $0.58 \pm 0.02$ &  0.010& 182.1 \\ 
 b&$\eennz$&  1&$ 1.30 \pm  0.17$ & $ 0.72 \pm  0.09$ & $ 0.58 \pm  0.15$ & $0.46 \pm 0.06$ &  0.013& 181.5 \\ 
 c&$\mmnnz$&  2&$ 0.97 \pm  0.15$ & $ 0.54 \pm  0.08$ & $ 0.44 \pm  0.13$ & $0.35 \pm 0.05$ &  0.013& 181.5 \\ 
 d&$\qqeez$&  5&$ 3.51 \pm  0.38$ & $ 2.79 \pm  0.13$ & $ 0.72 \pm  0.36$ & $0.69 \pm 0.03$ &  0.037& 174.7 \\ 
 e&$\bbeez$&  1&$ 0.68 \pm  0.07$ & $ 0.59 \pm  0.05$ & $ 0.09 \pm  0.06$ & $0.52 \pm 0.05$ &  0.010& 174.7 \\ 
 f&$\qqmmz$&  2&$ 3.22 \pm  0.12$ & $ 3.05 \pm  0.10$ & $ 0.18 \pm  0.08$ & $0.75 \pm 0.02$ &  0.037& 174.7 \\ 
 g&$\bbmmz$&  1&$ 0.73 \pm  0.04$ & $ 0.71 \pm  0.04$ & $ 0.02 \pm  0.02$ & $0.63 \pm 0.05$ &  0.010& 174.7 \\ 
 h&$\qqttz$&  1&$ 1.33 \pm  0.13$ & $ 1.14 \pm  0.10$ & $ 0.18 \pm  0.08$ & $0.28 \pm 0.03$ &  0.037& 174.7 \\ 
 i&$\bbttz$&  0&$ 0.30 \pm  0.06$ & $ 0.25 \pm  0.04$ & $ 0.04 \pm  0.04$ & $0.22 \pm 0.04$ &  0.010& 174.7 \\ 
 j&$\xbttz$&  1&$ 0.29 \pm  0.05$ & $ 0.27 \pm  0.04$ & $ 0.02 \pm  0.01$ & $0.24 \pm 0.04$ &  0.010& 174.7 \\ 
 k&$\qqnnz$& 20&$ 16.1 \pm   1.2$ & $  8.5 \pm   0.7$ & $ 7.51 \pm  0.94$ & $0.35 \pm 0.03$ &  0.219& 174.2 \\ 
 l&$\bbnnz$&  5&$ 2.18 \pm  0.24$ & $ 1.90 \pm  0.20$ & $ 0.28 \pm  0.14$ & $0.29 \pm 0.03$ &  0.061& 174.2 \\ 
 m&$\qqqqz$& 45&$ 50.1 \pm   4.0$ & $ 12.5 \pm   0.8$ & $ 37.6 \pm   3.9$ & $0.38 \pm 0.03$ &  0.300& 174.7 \\ 
 n&$\qqxbz$& 11&$ 6.74 \pm  0.75$ & $ 3.57 \pm  0.23$ & $ 3.16 \pm  0.72$ & $0.14 \pm 0.01$ &  0.235& 174.7 \\ 
 o&$\qqbbz$&  7&$ 7.87 \pm  0.64$ & $ 5.49 \pm  0.32$ & $ 2.38 \pm  0.56$ & $0.21 \pm 0.02$ &  0.235& 174.7 \\ 
\hline
\end{tabular}     
\end{center}
\caption[183 summary]{ 
Observed number of 
events, $\nobs$, the total
Standard Model expectation, $\smtot$,
the expected number of $\PZ$-pairs, $\nsm$,
background expectation, $\nback$, and
efficiencies, $\eff$, for the 183\,GeV and 189\,GeV data. 
$\br$ is the product branching
ratio for the final state
which is calculated directly from 
$\PZ$ resonance data~\cite{bib:pdg}.
Note that the efficiencies 
for selections with b-tags are given relative to 
the fraction of hadronic final states which
contain a $\PZ$ boson decaying to $\bbbar$.  
For selections of events with hadronic final states, but
without b-tags, the efficiencies are relative to
those hadronic final states which do not include
a $\PZ$ boson decaying to $\bbbar$.
The errors include contributions from the
common systematic errors.
\label{tab:zzxsec}
}
\end{table}

\begin{figure}
\begin{center}
    \mbox{\epsfxsize16cm\epsffile{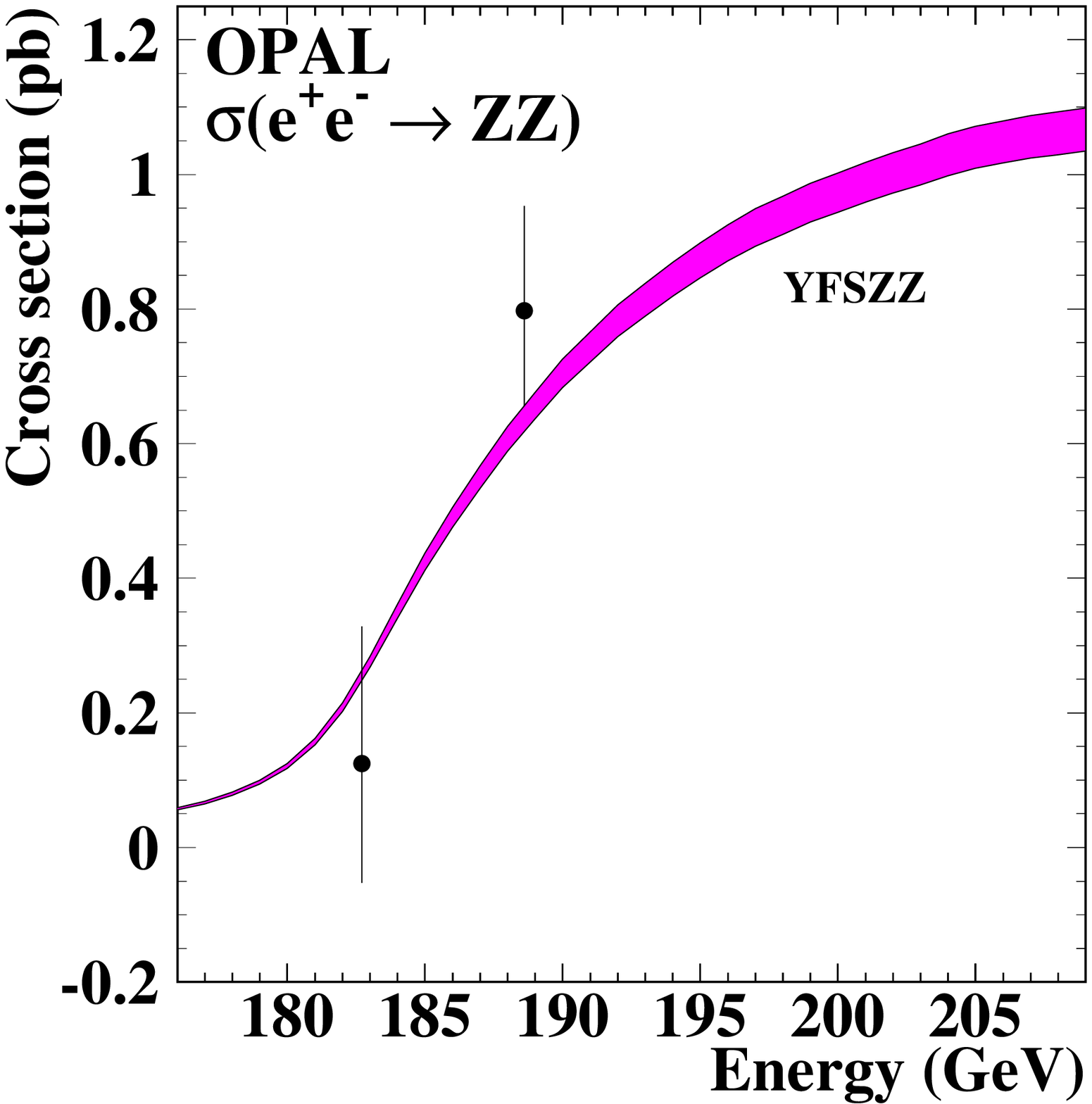}}
\end{center}
\caption[ZZ cross section]{
\label{fig:zzxsec}
The OPAL measurements
of the NC2 $\PZ$-pair production cross section.  
The shaded band shows the prediction,
using the default settings, 
of the YFSZZ Monte Carlo for the total cross section. The
band indicates a theoretical scale uncertainty
of $\pm 3$\%.
}
\end{figure}

The cross section at each energy is determined with
a maximum likelihood fit using a Poisson probability
density convolved with Gaussians to describe the
uncertainties on efficiencies and backgrounds.  
The expected number of events in each channel, $\nexp$,
as function of the $\PZ$-pair cross section, $\xsec$, 
is given by
\begin{equation}
\nexp  =  \xsec \lint \eff \br  + \nback.
\end{equation} 
The efficiencies, $\eff$,
include the effects of off-shell $\PZ$ bosons that 
are produced
outside of our kinematic acceptance.
Our main result, 
the NC2 $\PZ$-pair cross sections obtained from the
fits, is
$$
\begin{array}{lcll}
\xsec  (183 \ \mathrm{GeV})   & = & \xsca & \mathrm{pb} \\
&&& \\
\xsec  (189 \ \mathrm{GeV})  & = & \xscb & \mathrm{pb} .\\
\end{array}
$$
The 183\,GeV result corresponds to a 95\% C.L. upper limit
(normalized to the region $\xsec >0$)
of $\xsclma\ \pb$. 
The comparison of these measurements with the YFSZZ prediction,
using the coupling of $\PZ$ bosons to electrons measured
at the $\PZ$ resonance, is shown in Figure~\ref{fig:zzxsec}.  
The results are consistent with the YFSZZ prediction and with
the measurements presented in Reference~\cite{bib:l3}.

Our results assume that, apart from the backgrounds
discussed above,  only $\PZ\PZ$ production contributes inside
our kinematic region.  Possible effects of Higgs boson
production have been ignored.

In order to check the results for consistency with the 
expected fraction of $\bbbar$ final states,
we perform a second fit where the
branching ratio of the $\PZ$ boson to $\bbbar$
is a free parameter.  
The relative
branching ratios of the $\PZ$ to other fermion pairs
are fixed to their measured values.
The resulting branching ratio and cross section 
for the 189\,GeV data are
$\mbox{Br($\PZ \to \bbbar)$} = \bbrb $
and $ \xsec = \bbxs \mbox{pb}$.
The measured branching ratio 
can be compared with the world average
as measured at the $\PZ$ resonance of
$ \mbox{Br($\PZ \to \bbbar)$} = 0.1516 \pm 0.0009~$\cite{bib:pdg}.
The cross section 
measured without constraining 
the branching ratio is also consistent with
our main result and the YFSZZ prediction.
The smaller error on the main result cross section 
illustrates the advantage of classifying the hadronic systems
as $\bbbar$ or non-$\bbbar$.
The  183\,GeV data
sample is too small to extract a meaningful value of 
the $\PZ \to \bbbar$ branching ratio.

%
%

Limits on anomalous triple gauge couplings were set using
the total cross section and the
$|\cos \theta_{\mathrm{Z}}|$ distribution of our data.  
Here $\theta_{\mathrm{Z}}$ is 
the polar angle of the $\PZ$ bosons produced.
In this study we varied the real and imaginary
parts of the $\PZ\PZ\PZ$ and $\PZ\PZ\gamma$ 
anomalous couplings parameterized by the
form factors $\ZZZ{4}$,  $\ZZZ{5}$, 
$\ZZG{4}$ and $\ZZG{5}$ as
defined in Reference~\cite{bib:hagiwara} and
implemented in the YFSZZ Monte Carlo.
The real and imaginary parts of each coupling 
were varied separately with all others
fixed to zero.

For this study we consider the effect of the anomalous
couplings on the total cross section at $\roots = 183$\,GeV and
on the cross section in four bins of $| \cos \theta_{\mathrm{Z}}|$ at
$\roots = 189$\,GeV.
The selection efficiencies for all
final states are parameterized
as function of anomalous couplings.
At $\roots = 189$\,GeV the parameterization
is done separately for each bin in $| \cos \theta_{\mathrm{Z}} |$.
For values of the anomalous
couplings larger than unity, much of the production
of the final state fermions occurs at
$|\cos\theta| \simeq 1$ where 
the efficiency for most channels is reduced by a
factor of $\sim 0.5$.
An uncertainty of 10\%, dominated by Monte Carlo
statistical errors, is assigned to the
correction we apply to these efficiencies.

The 95\% C.L. limits on the anomalous couplings obtained
from the maximum likelihood fit are given
in Table~\ref{tab:anlim}.
With the exception of the $\ZZZR{5}$ coupling, 
the limits are insensitive to the sign and complex
phase of the couplings.

\begin{table}
\begin{center}
\begin{tabular}{|c|c|c|}
\hline
Coupling & 95\% C.L. Lower Limit &  95\% C.L. Upper Limit \\
\hline
 $ \ZZZR{4} $ & $ -2.1 $ & $    2.1$ \\
 $ \ZZZI{4} $ & $ -2.1 $ & $    2.1$ \\
 $ \ZZZR{5} $ & $ -6.2 $ & $    4.4$ \\
 $ \ZZZI{5} $ & $ -6.4 $ & $    6.4$ \\
 $ \ZZGR{4} $ & $ -1.2 $ & $    1.2$ \\
 $ \ZZGI{4} $ & $ -1.2 $ & $    1.2$ \\
 $ \ZZGR{5} $ & $ -3.9 $ & $    3.6$ \\
 $ \ZZGI{5} $ & $ -3.8 $ & $    3.9$ \\
\hline
\end{tabular}
\end{center}
\caption{
\label{tab:anlim} The 95\% confidence level
limits on possible anomalous triple gauge couplings.
}
\end{table}


\section{Conclusion} 

The production cross section
of $\eetozz$ has been measured using the final states
$\llll$, $\llnn$, $\qqll$, $\qqnn$, and  $\qqqq$.
The number of observed events, the
background expectation from Monte Carlo 
and the calculated efficiencies have been combined to measure
the production cross section of the process $\eetozz$.  
Our measured cross sections include the
effects of background and efficiency uncertainties.

We have determined the cross section for $\epem \to \PZ \PZ$ 
separately at average 
center-of-mass energies of $\emna \pm 0.05$\,GeV and 
$\emnb \pm 0.04$\,GeV.  
The NC2 $\PZ$-pair
cross sections were determined to be
$$
\begin{array}{lcll}
\xsec (183 \ \mathrm{GeV}) & = & \xsca & \mathrm{pb}\\
&&& \\
\xsec (189 \ \mathrm{GeV}) & = & \xscb & \mathrm{pb} .\\
\end{array} 
$$
At the lower center-of-mass energy, 
the 95\% C.L. upper limit on the cross section
is \xsclma\ pb.
The measurements at both energies are consistent
with the Standard Model expectations.
No evidence is found for anomalous neutral current triple gauge
couplings.  The 95\% confidence level limits are listed 
in Table~\ref{tab:anlim}.

\section*{Acknowledgements}
We particularly wish to thank the SL Division for the efficient operation
of the LEP accelerator at all energies
 and for their continuing close cooperation with
our experimental group.  We thank our colleagues from CEA, DAPNIA/SPP,
CE-Saclay for their efforts over the years on the time-of-flight and trigger
systems which we continue to use.  In addition to the support staff at our own
institutions we are pleased to acknowledge the  \\
Department of Energy, USA, \\
National Science Foundation, USA, \\
Particle Physics and Astronomy Research Council, UK, \\
Natural Sciences and Engineering Research Council, Canada, \\
Israel Science Foundation, administered by the Israel
Academy of Science and Humanities, \\
Minerva Gesellschaft, \\
Benoziyo Center for High Energy Physics,\\
Japanese Ministry of Education, Science and Culture (the
Monbusho) and a grant under the Monbusho International
Science Research Program,\\
Japanese Society for the Promotion of Science (JSPS),\\
German Israeli Bi-national Science Foundation (GIF), \\
Bundesministerium f\"ur Bildung, Wissenschaft,
Forschung und Technologie, Germany, \\
National Research Council of Canada, \\
Research Corporation, USA,\\
Hungarian Foundation for Scientific Research, OTKA T-029328, 
T023793 and OTKA F-023259.\\


\end{document}